\documentclass[]{spie}

\usepackage[utf8]{inputenc}

\usepackage{tikz}
\usepackage{pgfplots}
\pgfplotsset{compat=1.5}

\usepackage{amsmath,amsfonts,amssymb}

\usepackage{graphicx}
\usepackage[colorlinks=true, allcolors=blue]{hyperref}

\usepackage{txfonts}
\usepackage{textcomp}
\usepackage{eurosym}

\newcommand{\micron}{\mbox{$\mu$m}}
\renewcommand{\deg}{\mbox{deg}}

\title{COATLI: an all-sky robotic optical imager with 0.3 arcsec image quality}

\author[a]{Alan~M.~Watson}
\author[a]{Salvador~Cuevas~Cardona}
\author[a]{Luis~C.~Álvarez~Núñez}
\author[a]{Fernando~Ángeles}
\author[a]{Rosa~L.~Becerra-Godínez}
\author[a]{Oscar~Chapa}
\author[a]{Alejandro~S.~Farah}
\author[a]{Jorge~Fuentes-Fernández}
\author[b]{Liliana~Figueroa}
\author[a]{Rosalía~Langarica~Lebre}
\author[b]{Fernando~Quirós}
\author[b]{Carlos~G.~Román-Zúñiga}
\author[a]{Jaime~Ruíz-Diáz-Soto}
\author[b]{Carlos~G.~Tejada}
\author[a]{Silvio~J.~Tinoco}

\affil[a]{Instituto de Astronomía, Universidad Nacional Autónoma de México, Apartado Postal 70-264, 04510~México, México}
\affil[b]{Instituto de Astronomía, Universidad Nacional Autónoma de México, Apartado Postal 106, 22860~Ensenada, Baja California, México}

\authorinfo{Further author information: Send correspondence to A.M.W. E-mail: alan@astro.unam.mx}

\begin{document} 
\maketitle

\begin{abstract}
COATLI will provide 0.3 arcsec FWHM images from 550 to 900 nm over a large fraction of the sky. It consists of a robotic 50-cm telescope with a diffraction-limited fast-guiding imager. Since the telescope is small, fast guiding will provide diffraction-limited image quality over a field of at least 1 arcmin and with coverage of a large fraction of the sky, even in relatively poor seeing. The COATLI telescope will be installed at the at the Observatorio Astronómico Nacional in Sierra San Pedro Mártir, México, during 2016 and the diffraction-limited imager will follow in 2017.
\end{abstract}


\section{INTRODUCTION}

The Observatorio Astronómico Nacional (OAN) in Sierra San Pedro Mártir (SPM) has an excellent sky for astronomy. In particular, the seeing is similar to that at sites in Hawaii, the Andes, and the Canary Islands. Skidmore et al.\@\cite{skidmore-2009} measured the 25\%, 50\%, and 75\% values of the seeing at 500 nm to be 0.61, 0.79, and 1.12 arcsec FWHM with corresponding values of the Fried parameter $r_0$ of 16.9, 13.0, and 9.2 cm. Unfortunately, none of the existing telescopes or common-user instruments at the OAN/SPM are able to produce seeing-limited images; all are limited by dome seeing, static aberrations, guiding problems, and/or induced vibrations. Typically, the existing telescopes give images of 1.0 to 1.5 arcsec FWHM.

With the COATLI project, we aim to produce images from the OAN/SPM with FWHM of 0.25 to 0.35 from 550 nm to 920 nm (the $riz$ bands), better even the excellent seeing at SPM. Furthermore, we aim to do so routinely (i.e., even in relatively poor seeing), with excellent sky coverage, and with a robotic telescope. The one drawback of our approach is that we require a relatively small 50-cm telescope, which will limit us to science on relatively bright sources. We will achieve this image quality by using a combination of fast tilt correction (fast guiding or fast image stabilization) and active optics.

COATLI means “twin” in the indigenous Mexican Nahuatl language. This is a reference to the two channels and two detectors. Post facto, we twisted language to have make it refer to “Corrector de Óptica Áctiva y de Tilts al Límite de dIfracción” or “active optics and tilts corrector at the diffraction limit”.

\section{DIFFRACTION LIMITED IMAGES FROM FAST TILT CORRECTION}

In this section we will explain the basic idea that will allow COATLI to achieve diffraction-limited performance with a 50-cm telescope at wavelengths of 550--920 nm.

\subsection{Analytic Theory}

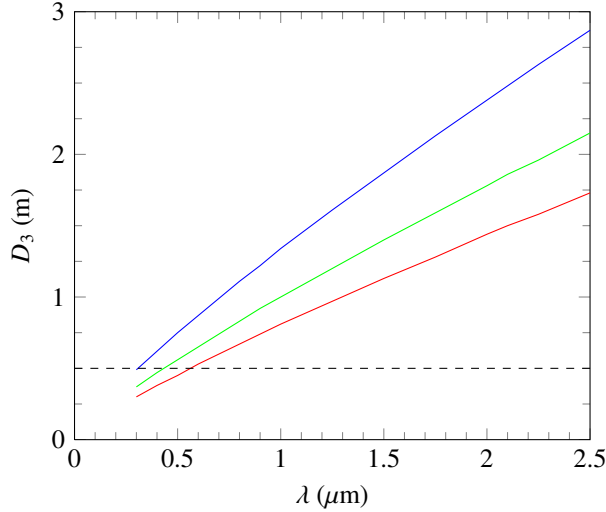
\begin{figure}
\begin{center}
\begin{tikzpicture}
\begin{axis}[
   xlabel={$\lambda$ (\micron)},
   ylabel={$D_3$ (m)},
   ymin=0,
   ymax=3,
   xmin=0,
   xmax=2.5,
   minor x tick num=4,
   minor y tick num=3,
]
\addplot[red] table[x index=0,y index=1]{figure-D3.tsv};
\addplot[green] table[x index=0,y index=2]{figure-D3.tsv};
\addplot[blue] table[x index=0,y index=3]{figure-D3.tsv};
\addplot[dashed]  table {
0.0 0.5
2.5 0.5
};
\end{axis}
\end{tikzpicture}
\end{center}
\caption{The value of $D_3$ as a function of wavelength in 25\% (blue), 50\% (green), and 75\% (red) conditions at SPM. We see that a 50-cm telescope (dashed line) gives good correction in the $r$ band (550--690 nm) even in 50\% conditions and in the $i$ band (690--820 nm) even in 75\% conditions.}
\label{figure:D3}
\end{figure}

Noll\cite{noll-1976} showed that the residual RMS phase variance after perfect tilt correction on a telescope of diameter $D$ is
\begin{eqnarray}
\Delta_3 &=& 0.134 \left(\frac{D}{r_0}\right)^{5/3}~\mathrm{rad}^2.\\
&=&0.134 \left(\frac{D}{r_0'}\right)^{5/3}\left(\frac{\lambda'}{\lambda}\right)^2~\mathrm{rad}^2.
\end{eqnarray}
In the second expression, we have scaled $r_0$ from a fiducial value $r_0'$ at a fiducial wavelength $\lambda'$ as $r_0 = r_0'(\lambda/\lambda')^{6/5}$. 
In the extended Marechal approximation, the Strehl ratio $S_3$ of the partially-corrected image is
\begin{eqnarray}
S_3 \approx e^{-\Delta_3}.
\label{eqn:marechal}
\end{eqnarray}
Empirically, diffraction-limited correction requires a phase variance of 1 $\mathrm{rad}^2$ or less (corresponding to a Strehl ratio of about $1/e$). From this, we can determine that the maximum diameter $D_3$ for diffraction-limited tilt correction is
\begin{eqnarray}
D_3&\approx&3.3\, r_0\\
&\approx&3.3\, r_0'\left(\frac{\lambda}{\lambda'}\right)^{6/5}
\end{eqnarray}
Figure~\ref{figure:D3} shows the value of $D_3$ for 25\%, 50\%, and 75\% conditions at SPM. One can see that at longer wavelengths and in better conditions, one can achieve diffraction-limited performance with a larger telescope. For example, in median conditions, one can see that diffraction-limited performance in $r$ (550--690 nm) in 50\% conditions can be achieved with a telescope no larger than about 50 cm whereas in $K$ (2.2 {\micron}) it can be achieved with a telescope of almost 2.0 m. 

We now change focus to consider the behavior of a telescope of fixed diameter $D$ at different wavelengths.
We can determine that the shortest wavelength $\lambda_3$ for diffraction-limited tilt correction is
\begin{eqnarray}
\lambda_3&\approx&0.37\, \lambda'\left(\frac{D}{r_0'}\right)^{5/6}
\end{eqnarray}
For a telescope of diameter $D_3$, or equivalently a telescope operating at wavelength $\lambda_3$, the FWHM of the diffraction-limited core is roughly
\begin{eqnarray}
\epsilon_3 \approx \frac{\lambda}{D_3} \approx 0.30 \frac{\lambda}{r_0} \approx 0.30 \epsilon_0,
\end{eqnarray}
in which $\epsilon_0 = \lambda/r_0$ is the seeing-limited FWHM for a large telescope. Thus, in this optimal case, images are significantly better than the seeing limit. On the other hand, we see that even in the optimal case, there is a limit to how good the images can be; they can never be better than about 1/3 of the seeing FWHM. We can also determine the wavelength $\lambda_0$ at which the diffraction-limit of the telescope is equal to the large-telescope seeing limit, 
\begin{eqnarray}
\lambda_0&\approx& \lambda'\left(\frac{D}{r_0'}\right)^{5/6}\\
&\approx&2.7\lambda_3.
\end{eqnarray}
We can now identify three wavelength regimes:
\begin{enumerate}
\item[(a)]$\lambda < \lambda_3$. The system will deliver seeing-limited images with Strehl ratios of less than 10\%, since pure tilt correction is unable to reduce the residual phase variance sufficiently to produce diffraction-limited images.


\item[(b)]$\lambda_3 \le \lambda < \lambda_0$. The system will produce diffraction-limited images, with FWHM narrower than the seeing FWHM. The FWHM will grow linearly with wavelength from about 1/3 of the seeing-limit at $\lambda_3$ to the seeing limit at $\lambda_0$. The Strehl ratio will grow super-linearly with wavelength from about 35\% at $\lambda_3$ to about 80\% at $2\lambda_3$ and finally to about 1 at $\lambda_0$.

\item[(c)]$\lambda_0 \le \lambda$. Diffraction-limited images, but now broadened by telescope diffraction to be wider than the seeing limit.

\end{enumerate}

Since the phase variance over a telescope of a fixed size decreases with wavelength, one can interpret regime (a) as having a phase variance that, even after removing the mean tilt, is larger than 1 $\mathrm{rad}^2$, regime (b) as having a phase variance that is intrinsically larger than 1 $\mathrm{rad}^2$, but after removing the mean tilt is less than 1 $\mathrm{rad}^2$, and regime (c) as having a phase variance that is intrinsically less than 1 $\mathrm{rad}^2$.

We can estimate the FWHM $\epsilon$ in regime (a) as the sum in quadrature of the large-telescope seeing limit $\epsilon_0$ and the diffraction-limit $\lambda/D$ and in regimes (b) and (c) as the diffraction-limit $\lambda/D$. We can estimate the Strehl ratio $S$ in regimes (b) and (c) using equation~(\ref{eqn:marechal}). 

For COATLI, a larger telescope with a HgCdTe infrared detector is ruled out for cost reasons. Instead, we have elected to use a 50-cm telescope with a red channel with a CCD detector operating from 550--920 nm (the Pan-STARRS $riz$ bands\cite{tonry-2012}) for science exposures and a blue channel with an EM CCD operating from 400--550 nm (the Pan-STARRS $g$ band\cite{tonry-2012}) mainly for tilt correction. The red channel normally operates in regime (b) and the blue channel normally operates in regime (a).

The expected behaviors of the FWHM $\epsilon$ and Strehl ratio $S$ for COATLI in 25\%, 50\%, and 75\% conditions are given in Figures~\ref{figure:epsilon} and \ref{figure:S}. We find that the theory predicts that with a 50 cm telescope we can achieve diffraction-limited performance to $\lambda_3 = 460$ nm in 25\% conditions, $\lambda_3 = 560$ nm in 50\% conditions, and $\lambda_3 = 770$ nm in 75\% conditions. Thus, even in 50\% conditions we should achieve diffraction-limited performance with a FWHM of about 0.25 arcsec in the $r$ filter (550--690 nm) and even in 75\% conditions we should achieve diffraction-limited performance with a FWHM of about 0.30 arcsec in the $i$ filter (690--820 nm).

\begin{figure}[p]
\begin{center}
\begin{tikzpicture}
\begin{axis}[
   xlabel={$\lambda$ (\micron)},
   ylabel={FWHM $\epsilon$ (arcsec)},
   ymin=0,
   ymax=1.5,
   xmin=0,
   xmax=1.5,
   minor x tick num=4,
   minor y tick num=4,
]
\addplot[blue,dotted] table[x index=0,y index=2]{figure-fwhm-S.tsv};
\addplot[green,dotted] table[x index=0,y index=3]{figure-fwhm-S.tsv};
\addplot[red,dotted] table[x index=0,y index=4]{figure-fwhm-S.tsv};
\addplot[blue,solid] table[x index=0,y index=5]{figure-fwhm-S.tsv};
\addplot[green,solid] table[x index=0,y index=6]{figure-fwhm-S.tsv};
\addplot[red,solid] table[x index=0,y index=7]{figure-fwhm-S.tsv};
\addplot[blue,dashed] table[x index=0,y index=1]{figure-yao-tilt.tsv};
\addplot[green,dashed] table[x index=0,y index=3]{figure-yao-tilt.tsv};
\addplot[red,dashed] table[x index=0,y index=5]{figure-yao-tilt.tsv};
\end{axis}
\end{tikzpicture}
\end{center}
\caption{The expected value of FWHM $\epsilon$ as a function of wavelength in 25\% (blue), 50\% (green), and 75\% (red) conditions at SPM. The solid lines are the analytic theory for COATLI with fast tilt correction and the dashed lines are the simulations. The dotted lines are the large-telescope seeing limit. The analytic theory clearly shows the three regimes described in the text: (a) poor correction; (b) diffraction-limited images better than the large-telescope seeing limit; and (c) diffraction-limited images worse than the large-telescope seeing limit.}
\label{figure:epsilon}

\begin{center}
\begin{tikzpicture}
\begin{axis}[
   xlabel={$\lambda$ (\micron)},
   ylabel={Strehl Ratio $S$},
   ymin=0,
   ymax=1,
   xmin=0,
   xmax=1.5,
   minor x tick num=4,
   minor y tick num=3,
]
\addplot[blue,solid] table[x index=0,y index=8]{figure-fwhm-S.tsv};
\addplot[green,solid] table[x index=0,y index=9]{figure-fwhm-S.tsv};
\addplot[red,solid] table[x index=0,y index=10]{figure-fwhm-S.tsv};
\addplot[blue,dashed] table[x index=0,y index=2]{figure-yao-tilt.tsv};
\addplot[green,dashed] table[x index=0,y index=4]{figure-yao-tilt.tsv};
\addplot[red,dashed] table[x index=0,y index=6]{figure-yao-tilt.tsv};
\end{axis}
\end{tikzpicture}
\end{center}
\caption{The expected value of the Strehl ratio $S$ as a function of wavelength for COATLI with fast tilt correction in 25\% (blue), 50\% (green), and 75\% (red) conditions at SPM. The solid lines are the analytic theory and the dashed lines are the simulations.}
\label{figure:S}
\end{figure}

\subsection{Simulations}

We performed simulations with François Rigaut’s Yao package\cite{rigaut-2002} to validate the analytic theory. We used a noiseless tilt sensor and 100 Hz correction. The results are shown in Figures~\ref{figure:epsilon} and \ref{figure:S}. We see that the simulations largely confirm the FWHM from the analytic theory, with a moderate additional broadening in poor seeing. However, the simulations suggest that the analytic theory overestimates the Strehl ratio by perhaps 10\%. We consider that the simulations to qualitatively and semi-quantitatively confirm the analytic theory.

\subsection{Experimental Confirmation}

\begin{figure}
\begin{center}
\includegraphics[width=0.3\linewidth]{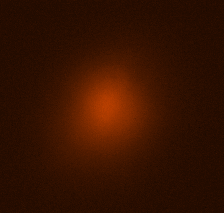}
\includegraphics[width=0.3\linewidth]{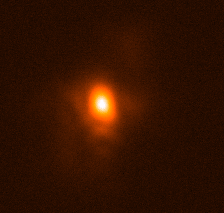}
\end{center}
\caption{Seeing-limited (left) and diffraction-limited (right) images of a star at 800 nm obtained with the OAXACA adaptive optics bench on the OAN Tonantzintla 1.0-meter telescope. The seeing was 2.0 arcsec and the aperture was 26 cm. The diffraction-limited image was obtained with fast tilt correction.}
\label{figure:oaxaca}
\end{figure}

We confirmed that tilt correction can produce diffraction-limited images with our experimental OAXACA adaptive optics bench on the 1.0-meter telescope of the OAN in Tonantzintla in central Mexico. With a seeing of 2.0 arcsec FWHM at 800 nm and an aperture of 26 cm, we obtained a diffraction-limited image with a FWHM of 0.63 arcsec and a Strehl ratio of 0.39, in good agreement with theory.\cite{becerra-2015} These images are shown in Figure~\ref{figure:oaxaca}. The value of $D/r_0$ for these conditions is about the same as expected for COATLI in 1.0 arcsec seeing.

\section{Science Cases}

Given the technological possibility of 0.25--0.35 arcsec images from 500--1000 nm, we asked our colleagues if they could identify science cases. We were given many examples:

\begin{itemize}
\item Eclipsing binaries in the Trapezium.
\item HII regions in nearby galaxies.
\item Galactic star clusters.
\item Location of gamma-ray bursts in their host galaxies.
\item Sub-stellar companions in the Solar neighborhood
\item Multiplicity in young clusters
\item Planetary nebulae in the bulge
\end{itemize}

These projects required imaging in broad-band $riz$ filters and narrow-band H$\alpha$. They are characterized by requiring good sky coverage and large samples or synoptic monitoring.

\section{Architecture}

On the basis of the considerations in the previous sections, we settled on the following architecture:

\begin{itemize}
\item
A 50-cm telescope. We chose an Astelco 50-cm Ritchey-Crétien telescope on a fast Astelco NTM-500 equatorial mount.
\item
An open enclosure. This is to minimize dome seeing. We elected to use an Astelco ARTS enclosure, which consists of a clamshell enclosure on a steel platform. This equipment is simple, robust, requires a minimum of maintenance, and has been proven in extreme conditions. The manufacturer guarantees it can open and close in a 90 km/h wind and survive a 180 km/h wind; in ten years of monitoring at the observatory the strongest gust we have seen is only 125 km/h.
\item
An elevated enclosure in a good site for seeing. The platform places the rotation axis of the mount about 3.9 meters above the platform feet. We will install the platform on 2.5 meter concrete columns to place the rotation axis of the mount about 6.4 meters above ground level. The mount itself will be installed on a massive 5.0 meter central column. We have also installed the platform on a rocky outcrop on the south-west side of the SPM summit (towards the prevailing wind) to have better seeing. These installations are shown in Figure~\ref{figure:installations}.

\begin{figure}
\begin{center}
\includegraphics[width=0.6\linewidth]{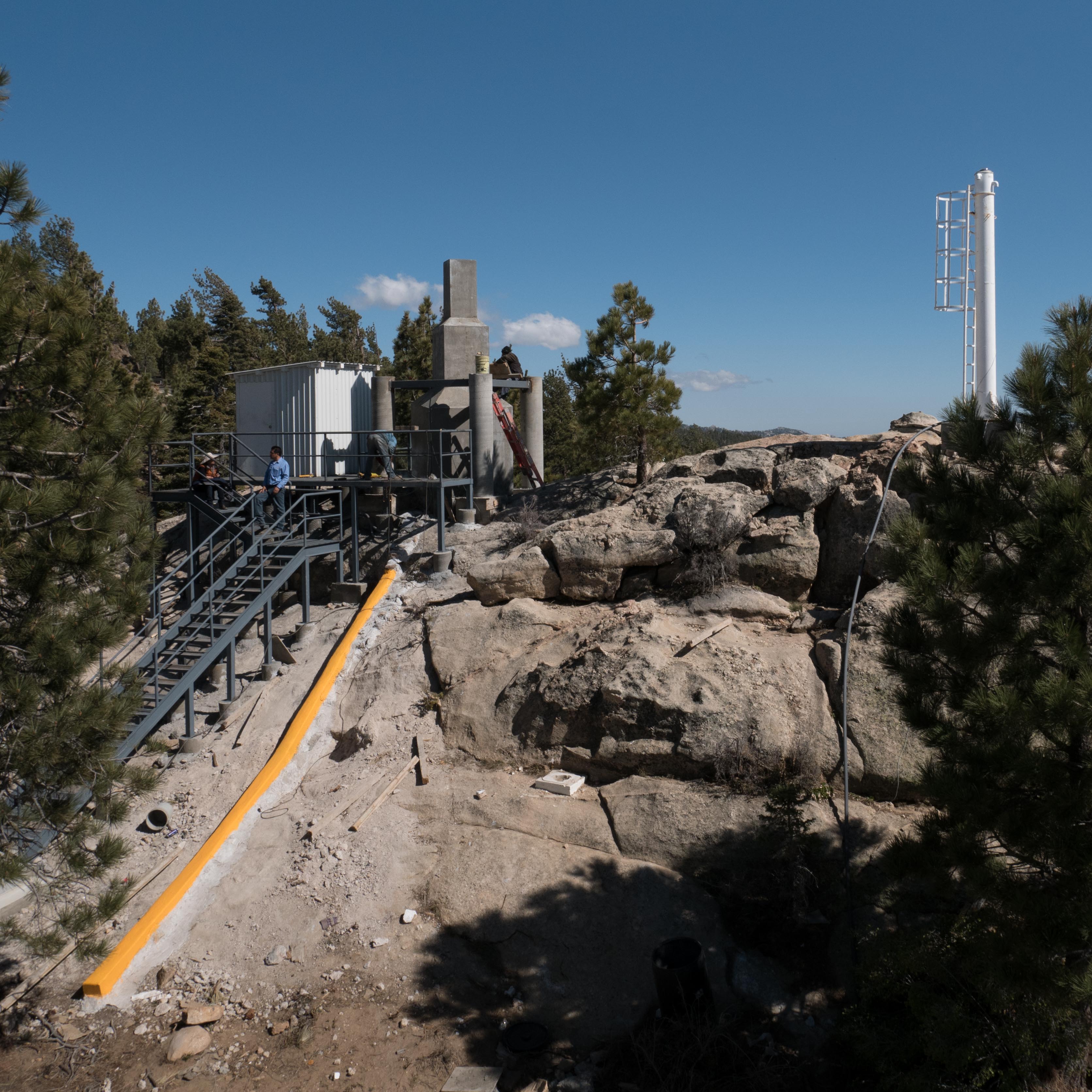}
\end{center}
\caption{The COATLI installations at SPM. The ARTS platform will be mounted on the four smaller columns (three of which are visible). The telescope and mount will be mounted on the massive central column. The shed to the left will contain the control computers and electronics. The installation is on a rocky outcrop to the south east of the existing 84-cm telescope. The white mast is an existing all-sky camera.}
\label{figure:installations}
\end{figure}

\item
A red channel operating from 550--920 nm with a conventional CCD. The chosen CCD is an Andor iKon-L 936 with a 4-stage TEC, BV finish, and $2048\times2048$ pixels. The scale is 0.12 arcsec/px, which gives critical sampling of the best images expected in the red, and the nominal field is 4.1 arcmin.

We initially considered a deep-depleted CCD, but rejected it for two reasons. First, none of our science cases required imaging in $y$ or beyond 920 nm. Second, current deep-depleted CCDs have significantly higher dark current than conventional CCDs, so reaching the low dark current specifications necessary for sensitive narrow-band imaging would require additional cooling and piping liquid coolant onto the telescope. We considered this was an unnecessary complication.
\item
A blue channel operating from 400--500 nm with an EM CCD for tilt correction. We chose an Andor iXon3 888 CCD with $1024\times1024$ pixels. The scale is 0.29 arcsec/pixel, which gives critical sampling of the best images expected in the blue, and the nominal field is 4.9 arcmin.
\item
A fast tilt mirror correcting both channels. We chose a Physik Instrumente S-325.30L piezo platform.
\item
A deformable mirror for active optics correction of the red channel, to maintain the telescope operating at the diffraction limit. We chose an Adaptica SATURN 48-element push-pull electrostatic mirror. The window of this mirror has poor transmission in the blue, so we elected to use it only in the red channel.
\item
A robotic control system based on our hardware and software for RATIR\cite{watson-2012} and OAXACA.
\end{itemize}

The detailed design of the instrument and telescope are described in the accompanying contributions by Fuentes-Fernández et al.\cite{fuentes-fernandez-2016} and Cuevas et al.\cite{cuevas-2016}

\section{Sky Coverage}

\subsection{Guide Star Availability}

\begin{table}
\caption{Sky Coverage}
\medskip
\label{table:sky-coverage}
\begin{center}
\begin{tabular}{lccc}
\hline
Case&Pessimistic&Neutral&Optimistic\\
\hline
$b = 20$ \deg&0.32&0.78&0.97\\
$b = 50$ \deg&0.17&0.53&0.82\\
$b = 90$ \deg&0.12&0.41&0.70\\
\hline
\end{tabular}
\end{center}
\end{table}

We simulated a realistic tilt sensor in Yao\cite{rigaut-2002} in order to determine the limiting guide star magnitude. In 50\% conditions, correction is essentially perfect to $g \approx 13$ and then falls away to a limiting magnitude of 16.

The sky coverage then depends on the isokinetic angle, which has not been extensively studied. Our requirements are actually quite relaxed compared to typical applications in adaptive optics. For example, an 8-meter telescope operating with an adaptive optics system giving diffraction limited performance in $K$ of 0.050 arcsec requires a guide star close enough that its relative movement is less than 0.025 arcsec. Since our images are at best 0.25 arcsec, our requirements are 5 times more lax at 0.125 arcsec.

Anecdotal evidence from other telescopes\cite{tonry-1997,jim-2000} suggests that the isokinetic angle for this type of correction is up to 3 arcmin. We will define three cases: (a) “pessimistic” with an isokinetic angle of 1 arcmin, (b) “neutral” with an angle of 2 arcmin, and (c) “optimistic” with an angle of 3 arcmin. Combining our simulations for the Strehl ratio as a function of guide star magnitude and a star density model\cite{bahcall-1980} we have determine the probability of finding a sufficiently close and sufficiently bright guide star for diffraction-limited correction as a function of galactic latitude $b$. These values are given in Table~\ref{table:sky-coverage}. In the neutral case, the sky coverage ranges from 78\% near the plane ($b = 20$ \deg) to 41\% in the pole ($b = 90$ \deg). In the optimistic and pessimistic cases, the sky coverage is obviously better and worse. These figures are sufficiently high compared to typical natural guide star adaptive optics systems that we stretch them to suggest that COATLI is (nearly) an all-sky system.

\subsection{Seeing Away from the Zenith}

The seeing degrades away from the zenith as $\epsilon_0 \propto X^{3/5}$ or $r_0 \propto X^{-3/5}$, in which $X$ is the airmass, the secant of the zenith distance.  This causes worse correction away from the zenith. For example, consider 50\% conditions with $r_0 = 13$ cm at the zenith. At a zenith distance of 56 {\deg}, the seeing has degraded to $r_0 = 9.2$ cm, equal to 75\% conditions at the zenith.

\subsection{Atmospheric Dispersion}

COATLI does not have an atmospheric dispersion corrector. This is mainly a problem for the $r$ band, since at a zenith distance of 35 {\deg} differential atmospheric refraction will be equal to the diameter of the diffraction-limited core. For the $iz$ and narrow-band filters, this will not be a problem until a zenith distance of at least 60 {\deg}, by which point the effective seeing is poor anyway.

\section{Sensitivity}

For aperture photometry of the diffraction-limited core in 50\% conditions at the zenith, we estimate 10$\sigma$ limiting magnitudes in dark/bright time of 23.8/23.4 in $r$, 23.3/22.9 in $i$, and 22.6/22.3 in $z$ in 1000 seconds.

\section{Status}

The total cost of COATLI, including the instrument, telescope, enclosure, mount, and installations, is about US\$500,000. COATLI is fully funded. 

COATLI passed its initial design review in July 2014. The telescope and installations passed their final design review in April 2015. We expect to present the final design review of the instrument in August 2016.

The installations were originally scheduled to be finished in October 2015. However, the local contractor finally delivered them in May 2016. We plan to install the telescope in September 2016, after the summer lightning season. 

We will initially operate the telescope with an interim imager consisting of an Finger Lakes Instrument ML3200ME with a Kodak KAF-3200ME CCD. This CCD has $2184 \times 1510$ pixels each $6.8$ {\micron} square, which corresponds to a field of $12.7 \times 8.8$ arcmin with 0.35 arcsec pixels. Thus, the interim images will be well-suited to seeing-limited observations. The interim imager will have a filter wheel with $BVRI$ and clear filters.

We expect to commission the definitive diffraction-limited imager in 2017.

\acknowledgements

Science cases were contributed by Nat Butler, Rafael Costero, Juan José Downes, Alexander Kutyrev, William Lee, Michael Richer, Carlos Román-Zúñiga, Sebastián Sánchez, and Alan Watson. COATLI has been funded by the Instituto de Astronomía of the UNAM, CONACyT proposals 232649, 260369, and 271117 (Laboratorio Nacional de Astrofísica en San Pedro Mártir), and UNAM/PAPIIT projects IG100414 and IT102715.

\bibliography{coatli} 
\bibliographystyle{spiebib} 

\end{document}